\documentstyle[12pt]{article}
\newcommand{\eq}{\begin{equation}}
\newcommand{\en}{\end{equation}}
\def\opensquare{\thicklines\raisebox{-8pt}{\framebox(20,20){}}}
\def\sector#1#2{\ {\scriptstyle #1}\hspace{1mm}
\mathop{\opensquare}\limits_{\raisebox{-1mm}{$\scriptstyle#2$}}\hspace{1mm}}

\begin{document}
\addtolength{\baselineskip}{.3mm}
\thispagestyle{empty}
\begin{center}
{\large\sc Three generation Distler-Kachru models}\\[13mm]
{\sc Yuri M. Malyuta}\\[3mm]
{\it Institute for Nuclear Research National Academy of Sciences of Ukraine,\\
 252028 Kiev, Ukraine}\\[7mm]
{\sc Nikolay N. Aksenov}\\[3mm]
{\it Glushkov Institute
of Cybernetics   National Academy of Sciences of Ukraine,\\
40 Acad. Glushkov Prsp.
252650 MSP, Kiev 22,
Ukraine}\\
E-mail: aks@d310.icyb.kiev.ua\\[18mm]
\end{center}
\vspace{1cm}
\begin{center}
{\sc Abstract}
\end{center}
\noindent
 Distler-Kachru models which yield three generations of chiral fermions
with gauge group $SO(10)$ are found. These models have mirror
partners.
\newpage

\section{Introduction}
 Distler--Kachru models \cite{DK}\ are special  cases of
(0,2) Landau-Ginzburg orbifolds \cite{KW}. To construct these models
methods of algebraic geometry are being used. These models make it possible
to determine the number of generations of chiral fermions. For this
purpose it suffices to compute the Euler characteristic of the corresponding
vector bundle.

In the present work we use a practical method of computing
the Euler characteristic suggested by Kawai and Mohri \cite{KM}\ for
a large class of Distler--Kachru models. By means of this method we provide
some examples of models which yield three generations  of chiral fermions
with gauge group $SO(10)$. These models have mirror partners \cite{YU}.
\section{A class of Distler--Kachru models}
The Lagrangian for Distler--Kachru models \cite{KM} is given by
\eq
{\cal L}=2i\int d\theta^{+}d\overline{\theta}^{+}
\Phi^{\overline{i}}\partial_{-}\Phi^{i}
-\int d\theta^{+}d\overline{\theta}^{+}\Lambda^{k}
\Lambda^{\overline{k}}
+\int d\theta^{+}F_{k}\Lambda^{k}+
\int d\overline{\theta}^{+}F_{\overline{k}}
\Lambda^{\overline{k}},
\label{1}
\en
where bosonic  superfields
$\Phi^{i}$ and fermionic ones $\Lambda^{k}$ have the component
expansions
\begin{eqnarray*}
  \begin{array}{lll}
\Phi^{i}&=&\phi^{i}+\theta^+ \psi^i+i\;\overline{\theta}^+\theta^+
\partial_+\phi^i\;,\qquad 1\ll i\ll N\;,\\
\Lambda^{k}&=&\lambda^k-\theta^+ l^k +i\;\overline{\theta}^+\theta^+
\partial_+\lambda^k\;,\qquad 1\ll k\ll M\;,
  \end{array}
\end{eqnarray*}
and superpotentials $F_k$ are homogeneous polynomials of
$\Phi^{i}$
satisfying
\[
F_{k}(x^{\omega_{1}}\Phi^{1},x^{\omega_{2}}\Phi^{2},...,
x^{\omega_{N}}\Phi^{N})=x^{1-\rho_{k}}
F_{k}(\Phi^{1},\Phi^{2},...,\Phi^{N}),
\]
with $\omega_i$ and $\rho_k$ being rational numbers.

Following \cite{KM}, let us reformulate superconformal model (\ref{1})
into the language of algebraic geometry. Suppose that $X$ is a
 $D=N-M+r$ dimensional  complete
intersection defined by
\[
  X=\{ p\in {\bf WP}^{N-1}_{{\rm w}_1,\ldots,{\rm w}_N}
\mid F_{r+2}(p)=\cdots=F_{M}(p)=0\},
\]
where $F_{r+1+j}$ is a degree $d_j$ polynomial in the
coordinates  of
the weighted projective space ${\bf WP}^{N-1}_{{\rm w}_1,\ldots,{\rm w}_N}$.
Let $E$ be the stable rank $r$ vector bundle over $X$ defined by the following
exact sequence
\[
0\rightarrow E\rightarrow \mathop{\bigoplus}\limits_{a=1}^{r+1}{\cal O}(n_a)\;
\mathop{\longrightarrow} {\cal O}(m)\rightarrow 0,
\]
where the  $n_a$ and $m$ being positive integers. Here $r=3,4,5$ yields
gauge group $E_6$, $SO(10)$, $SU(5)$. The vector bundle $E$ and tangent
bundle $T$ of $X$ must satisfy constraints on their Chern classes
\begin{eqnarray}
  c_1(T)=0\,,\nonumber\\
c_1(E)=0\,,\label{2}\\
 c_2(T)=c_2(E)\,,\nonumber
\end{eqnarray}
which are the anomaly cancellaton conditions. Conditions (\ref{2})
are tantamount to the system of Diophantine equations
\begin{eqnarray}
  \label{3}
\sum_i{\rm w}_i-\sum_j d_j&=&0\,,\nonumber\\
\sum_a n_a-m&=&0\,, \\
\sum_i{{\rm w}_i}^2-\sum_jd_j^2&=&\sum_a n_a^2-m^2\,. \nonumber
\end{eqnarray}
Bosonic  superfields
$\Phi^{i}$ can be interpreted as coordinates of $X$
and fermionic ones $\Lambda^{k}$ as sections of $E$.
\section{Orbifoldized elliptic genus}
To compute the  Euler characteristic $\chi(E)$, it is convenient
to start from orbifoldized elliptic genus \cite{KM}
\[
Z(\tau,z)=\frac{1}{m}\sum_{\alpha,\beta=0}^{m-1}
(-1)^{r(\alpha+\beta+\alpha\beta)}\sector\beta\alpha(\tau,z)\,,
\]
where
\[
\  \sector{\beta}{\alpha} (\tau,z)=
(-1)^{r\alpha\beta}e^{\pi ir(\alpha^2\tau+2\alpha z)}
\sector{0}{0}(\tau,z+\alpha\tau+\beta)\,,
\]
with
\[
\sector{0}{0}(\tau,z)=\frac{\prod_{k=1}^M P(\tau,(1-\rho_k)z)}
               {\prod_{i=1}^N P(\tau,\omega_i z)}\,,
\]
\[
P(\tau,z)=\vartheta_1(\tau,z)/\eta(\tau)\,,
\]
$\vartheta_1(\tau,z)$ is the Jacobi theta function, $\eta(\tau)$ is the
Dedekind eta function.

The modular and double quasi-periodicity properties of the
orbifoldized elliptic genus lead to the following relations
\[
  \begin{array}{ll}
&\displaystyle(\omega_1,\ldots,\omega_N)= \frac{1}{m}({\rm w}_1,
\ldots,{\rm w}_N)\,,\\[3mm]
&\displaystyle(\rho_1,\ldots,\rho_M)=\frac{1}{m}(n_1,\ldots,n_{r+1},
m-d_1,\ldots,m-d_{M-r-1})\,.
\end{array}
\]

An expansion
\[
  Z(\tau,z)=i^{M-N}q^{\frac{M-N}{12}}y^{-r/2}
\left[\chi_y(E)+{\rm O}(q)\right]
\]
connects the orbifoldized elliptic genus with the Hirzebruch genus
\[
  \chi_y(E)=\sum^{m-1}_{\alpha=0}\chi_{y}^{(\alpha)},
\]
 which consist of contributions from twisted sectors
\begin{eqnarray}
&&\chi_{y}^{(\alpha)}\!=\!
(-1)^{r\alpha}\!\left.\frac{\prod_{k=1}^M(-1)^{[\alpha\nu_k]}\Big[y^{\nu_{k}}
q^{\frac{\{\alpha\nu_{k}\}-1}{2}}\Big]^{\{\alpha\nu_{k}\}}
(1-y^{\nu_{k}}q^{\{\alpha\nu_{k}\}})
(1-y^{-\nu_{k}}q^{1-\{\alpha\nu_{k}\}})}
{\prod_{i=1}^N (-1)^{[\alpha\omega_i]}
 \Big[y^{\omega_{i}}q^{\frac{\{\alpha\omega_{i}\}-1}{2}}
\Big]^{\{\alpha\omega_{i}\}}
(1-y^{\omega_{i}}q^{\{\alpha\omega_{i}\}} )
(1-y^{-\omega_{i}}q^{1-\{\alpha\omega_{i}\}})}\,\right|_*,
\nonumber\\[2mm]
&&
\label{4}
\end{eqnarray}
where $y=e^{2\pi iz}$, $q=e^{2\pi i\tau}$, $\nu_k=1-\rho_k$, $\{x\}=x-[x]$,
$[x]$ denotes the greatest integer less than $x$, $|_*$ means that
we extract only  terms of the form $q^0y^{\rm integer}$ in the
expansion (\ref{4}).
\section{$SO(10)$ models}
We now use the formula (\ref{4}) to construct some three
generation models with $r=4$, $j=1$, $D=3$. The Euler characteristic
for these models is given by
\begin{equation}
  \chi(E)=-\chi_{y}(E)/y(y+1)(y-1).
\label{5}
\end{equation}

Consider the models defined by the following quantum numbers
\begin{eqnarray}
&&({\rm w}_1,\ldots,{\rm w}_5\,;d_1)=(n_1,\ldots,n_5\,;m)\,:\nonumber\\
&&(3,4,6,13,13;39)\,,\nonumber\\
&&(5,8,9,11,12;45)\,,\label{anti}\\
&&(10,12,13,15,25;75)\,.\nonumber
\end{eqnarray}
The system of equations (\ref{3}) are satisfied identically
by the data (\ref{anti}). The quantum numbers (\ref{anti})
(and (\ref{normal}) below)
had been used
by Klemm and  Schimmrigk \cite{KS} to obtain the three generation
(2,2) superconformal models. But whereas in the (2,2) case
the gauge group is $E_6$, in the  (0,2) case the gauge group is
$SO(10)$ for $r=4$. In tables 1--3 we display the contributions
from twisted sectors into the Hirzebruch genera of models (\ref{anti}).

\begin{center}
Table 1\\*[2mm]
{\small
\begin{tabular}{|r|c|r|c|r|c|}\hline
\multicolumn{6}{|c|}{$({\rm w}_i;d_1)=(n_a;m)=(3,4,6,13,13;39)$}\\ \hline
$\alpha$&$\chi_y^{(\alpha)}$
&$\alpha$&$\chi_y^{(\alpha)}$
&$\alpha$&$\chi_y^{(\alpha)}$\\ \hline
$0$&$1+25y-25y^{3}-y^{4} $
&$13$&$-6y^2 + 6y^3 $
&$26$&$-6y + 6y^2 $\\
$1$&$-y^3+y^{4} $
&$14$&$-y^2 + y^3 $
&$27$&$-2y^2 + 2y^3 $\\
$2$&$- y^2 + y^3 $
&$15$&$-2y^2 + 2y^3 $
&$28$&$-y^2 + y^3 $\\
$3$&$-2y^2 + 2y^3 $
&$16$&$-y^2 + y^3 $
&$29$&$-y + y^2 $\\
$4$&$-y^2 + y^3 $
&$17$&$-y + y^2 $
&$30$&$-2y^2 + 2y^3 $\\
$5$&$-y - y^2 $
&$18$&$-2y + 2y^2 $
&$31$&$-y^2 + y^3 $\\
$6$&$-2y + 2y^2 $
&$19$&$-y + y^2 $
&$32$&$-y + y^2 $\\
$7$&$-y^2 + y^3 $
&$20$&$-y^2 + y^3 $
&$33$&$-2y^2 + 2y^3 $\\
$8$&$-y + y^2 $
&$21$&$-2y^2 + 2y^3 $
&$34$&$-y^2 + y^3 $\\
$9$&$-2y + 2y^2 $
&$22$&$-y^2 + y^3 $
&$35$&$-y + y^2 $\\
$10$&$-y^2 + y^3 $
&$23$&$-y + y^2 $
&$36$&$-2y + 2y^2 $\\
$11$&$-y + y^2 $
&$24$&$-2y + 2y^2 $
&$37$&$-y + y^2 $\\
$$12&$-2y + 2y^2 $
&$$25&$-y + y^2 $
&$$38&$-1 + y $\\\hline
\hline
\multicolumn{6}{|c|}{$\chi_y(E)=-3y(1+y)(1-y)$}\\ \hline
    \end{tabular}}
\end{center}

\begin{center}
Table 2\\*[2mm]
{\small
\begin{tabular}{|r|c|r|c|r|c|}\hline
\multicolumn{6}{|c|}{$({\rm w}_i;d_1)=(n_a;m)=(5,8,9,11,12;45)$}\\ \hline
$\alpha$&$\chi_y^{(\alpha)}$
&$\alpha$&$\chi_y^{(\alpha)}$
&$\alpha$&$\chi_y^{(\alpha)}$\\ \hline
$ 0$&$1 + 12y - 12y^3- y^4$    &$15$&$ - y + y^2  $ &$30$&$  - y^2 + y^3 $\\
$ 1$&$-y^3 + y^4          $    &$16$&$ -y + y^2   $ &$31$&$ -y^2 + y^3   $\\
$ 2$&$-y^2 + y^3          $    &$17$&$ -y^2 + y^3 $ &$32$&$ -y + y^2     $\\
$ 3$&$-y + y^2            $    &$18$&$ 0          $ &$33$&$ -y + y^2     $\\
$ 4$&$-y + y^2            $    &$19$&$ -y^2 + y^3 $ &$34$&$ -y^2 + y^3   $\\
$ 5$&$0                   $    &$20$&$ 0          $ &$35$&$ 0            $\\
$ 6$&$-y^2 + y^3          $    &$21$&$ -y^2 + y^3 $ &$36$&$ 0            $\\
$ 7$&$-y + y^2            $    &$22$&$ -y + y^2   $ &$37$&$ -y^2 + y^3   $\\
$ 8$&$-y + y^2            $    &$23$&$ -y^2 + y^3 $ &$38$&$ -y^2 + y^3   $\\
$ 9$&$0                   $    &$24$&$ -y + y^2   $ &$39$&$ -y + y^2     $\\
$10$&$0                   $    &$25$&$ 0          $ &$40$&$ 0            $\\
$11$&$-y + y^2            $    &$26$&$ -y + y^2   $ &$41$&$ -y^2 + y^3   $\\
$12$&$-y^2 + y^3          $    &$27$&$ 0          $ &$42$&$ -y^2 + y^3   $\\
$13$&$-y^2 + y^3          $    &$28$&$ -y + y^2   $ &$43$&$ -y + y^2     $\\
$14$&$-y + y^2 $ &$29$&$ -y^2 + y^3 $ &$44$&$ -1 + y $\\\hline
\hline
\multicolumn{6}{|c|}{$\chi_y(E)=-3y(1+y)(1-y)$}\\ \hline
    \end{tabular}}
\end{center}

\begin{center}
Table 3\\*[2mm]
{\small
\begin{tabular}{|r|c|r|c|r|c|}\hline
\multicolumn{6}{|c|}{$({\rm w}_i;d_1)=(n_a;m)=(10,12,13,15,25;75)$}\\ \hline
$\alpha$&$\chi_y^{(\alpha)}$
&$\alpha$&$\chi_y^{(\alpha)}$
&$\alpha$&$\chi_y^{(\alpha)}$\\ \hline
$ 0$&$1 + 8y - 8y^3- y^4  $&$25$&$ - y^2 + y^3  $&$50$&$   - y +y^2    $\\
$ 1$&$-y^3 + y^4          $&$26$&$ -y^2 + y^3   $&$51$&$   0           $\\
$ 2$&$-y^2 + y^3          $&$27$&$ 0            $&$52$&$   -y^2 + y^3  $\\
$ 3$&$0                   $&$28$&$ -y + y^2     $&$53$&$   -y^2 + y^3  $\\
$ 4$&$-y + y^2            $&$29$&$ -y + y^2     $&$54$&$   0           $\\
$ 5$&$0                   $&$30$&$  2y - 2y^3   $&$55$&$   0           $\\
$ 6$&$0                   $&$31$&$ -y^2 + y^3   $&$56$&$   -y + y^2    $\\
$ 7$&$-y^2 + y^3          $&$32$&$ -y^2 + y^3   $&$57$&$   0           $\\
$ 8$&$-y^2 + y^3          $&$33$&$ 0            $&$58$&$   -y^2 + y^3  $\\
$ 9$&$0                   $&$34$&$ -y + y^2     $&$59$&$   -y + y^2    $\\
$10$&$0                   $&$35$&$ 0            $&$60$&$    2y -2y^3   $\\
$11$&$-y + y^2            $&$36$&$ 0            $&$61$&$   -y^2 + y^3  $\\
$12$&$0                   $&$37$&$ -y + y^2     $&$62$&$   -y + y^2    $\\
$13$&$-y^2 + y^3          $&$38$&$ -y^2 + y^3   $&$63$&$   0           $\\
$14$&$-y + y^2            $&$39$&$ 0            $&$64$&$   -y^2 + y^3  $\\
$15$&$2y  - 2y^3          $&$40$&$ 0            $&$65$&$   0           $\\
$16$&$-y^2 + y^3          $&$41$&$ -y^2 + y^3   $&$66$&$   0           $\\
$17$&$-y + y^2            $&$42$&$ 0            $&$67$&$   -y + y^2    $\\
$18$&$0                   $&$43$&$ -y + y^2     $&$68$&$   -y + y^2    $\\
$19$&$-y^2 + y^3          $&$44$&$ -y + y^2     $&$69$&$   0           $\\
$20$&$0                   $&$45$&$  2y - 2y^3   $&$70$&$   0           $\\
$21$&$0                   $&$46$&$ -y^2 + y^3   $&$71$&$   -y^2 + y^3  $\\
$22$&$-y + y^2            $&$47$&$ -y^2 + y^3   $&$72$&$   0           $\\
$23$&$-y + y^2            $&$48$&$ 0            $&$73$&$   -y + y^2    $\\
$24$&$0 $&$49$&$ -y + y^2 $&$74$&$ -1 + y $\\\hline
\hline
\multicolumn{6}{|c|}{$\chi_y(E)=-3y(1+y)(1-y)$}\\ \hline
    \end{tabular}}
\end{center}
{}From the results of tables 1--3 and formula (\ref{5}) it is clear
that $\chi(E)=3$. Hence, these models yield three
generation of chiral fermions
with $SO(10)$ gauge group.

The models (\ref{anti}) have mirror partners defined by the quantum numbers
\begin{eqnarray}
&({\rm w}_1,\ldots,{\rm w}_5\,;d_1)=(n_1,\ldots,n_5\,;m)\,:\nonumber\\
&(3,4,12,17,19;55)\,,\nonumber\\
&(4,7,9,10,15;45)\,,\label{normal}\\
&(4,4,5,5,7;25)\,.\nonumber
\end{eqnarray}
For these mirror partners $\chi(E)=-3$. The contributions from
twisted sectors into the Hirzebruch genera of mirror partners
(\ref{normal}) are shown in tables 4--6.

\begin{center}
Table 4\\*[2mm]
{\small
\begin{tabular}{|r|c|r|c|r|c|}\hline
\multicolumn{6}{|c|}{$({\rm w}_i;d_1)=(n_a;m)=(3,4,12,17,19;55)$}\\ \hline
$\alpha$&$\chi_y^{(\alpha)}$
&$\alpha$&$\chi_y^{(\alpha)}$
&$\alpha$&$\chi_y^{(\alpha)}$\\ \hline
$ 0$&$ 1 + 28y - 28y^3 - y^4   $&$19$&$-y^2 + y^3   $&$38$&$-y^2 + y^3   $\\
$ 1$&$ -y^3 + y^4              $&$20$&$-y^2 + y^3   $&$39$&$-y^2 + y^3   $\\
$ 2$&$ -y^2 + y^3              $&$21$&$-y^2 + y^3   $&$40$&$-y + y^2     $\\
$ 3$&$ -y^2 + y^3              $&$22$&$-y + y^2     $&$41$&$-y + y^2     $\\
$ 4$&$ -y^2 + y^3              $&$23$&$-y^2 + y^3   $&$42$&$-y^2 + y^3   $\\
$ 5$&$ -y^2 + y^3              $&$24$&$-y^2 + y^3   $&$43$&$-y^2 + y^3   $\\
$ 6$&$ -y^2 + y^3              $&$25$&$-y + y^2     $&$44$&$-y^2 + y^3   $\\
$ 7$&$ -y^2 + y^3              $&$26$&$-y + y^2     $&$45$&$-y + y^2     $\\
$ 8$&$ -y + y^2                $&$27$&$-y + y^2     $&$46$&$-y^2 + y^3   $\\
$ 9$&$ -y + y^2                $&$28$&$-y^2 + y^3   $&$47$&$-y^2 + y^3   $\\
$10$&$ -y^2 + y^3              $&$29$&$-y^2 + y^3   $&$48$&$-y + y^2     $\\
$11$&$ -y + y^2                $&$30$&$-y^2 + y^3   $&$49$&$-y + y^2     $\\
$12$&$ -y + y^2                $&$31$&$-y + y^2     $&$50$&$-y + y^2     $\\
$13$&$ -y + y^2                $&$32$&$-y + y^2     $&$51$&$-y + y^2     $\\
$14$&$ -y^2 + y^3              $&$33$&$-y^2 + y^3   $&$52$&$-y + y^2     $\\
$15$&$ -y^2 + y^3              $&$34$&$-y + y^2     $&$53$&$-y + y^2     $\\
$16$&$ -y + y^2                $&$35$&$-y + y^2     $&$54$&$-1 + y       $\\
$17$&$ -y + y^2                $&$36$&$-y + y^2     $&$  $&$             $\\
$18$&$ -y + y^2 $&$37$&$-y^2 + y^3 $&$ $&$ $\\\hline
\hline
\multicolumn{6}{|c|}{$\chi_y(E)=3y(1+y)(1-y)$}\\ \hline
    \end{tabular}}
\end{center}

\begin{center}
Table 5\\*[2mm]
{\small
\begin{tabular}{|r|c|r|c|r|c|}\hline
\multicolumn{6}{|c|}{$({\rm w}_i;d_1)=(n_a;m)=(4,7,9,10,15;45)$}\\ \hline
$\alpha$&$\chi_y^{(\alpha)}$
&$\alpha$&$\chi_y^{(\alpha)}$
&$\alpha$&$\chi_y^{(\alpha)}$\\ \hline
$ 0$&$ 1 + 15y - 15y^3 - y^4  $    &$15$&$ 0          $ &$30$&$ 0 $\\
$ 1$&$ -y^3 + y^4             $    &$16$&$ -y^2 + y^3 $ &$31$&$ -y + y^2 $\\
$ 2$&$ -y^2 + y^3             $    &$17$&$ -y + y^2   $ &$32$&$ -y + y^2 $\\
$ 3$&$ 0                      $    &$18$&$ - y + y^2  $ &$33$&$ 0 $\\
$ 4$&$ -y + y^2               $    &$19$&$ -y + y^2   $ &$34$&$ -y^2 + y^3$\\
$ 5$&$ 0                      $    &$20$&$ 0          $ &$35$&$ 0 $\\
$ 6$&$ 0                      $    &$21$&$ 0          $ &$36$&$ -y^2 + y^3$\\
$ 7$&$ -y^2 + y^3             $    &$22$&$ -y + y^2   $ &$37$&$ -y^2 + y^3$\\
$ 8$&$ -y + y^2               $    &$23$&$ -y^2 + y^3 $ &$38$&$ -y + y^2 $\\
$ 9$&$ - y + y^2              $    &$24$&$ 0          $ &$39$&$ 0 $\\
$10$&$ 0                      $    &$25$&$ 0          $ &$40$&$ 0 $\\
$11$&$ -y + y^2               $    &$26$&$ -y^2 + y^3 $ &$41$&$ -y^2 + y^3 $\\
$12$&$ 0                      $    &$27$&$ - y^2 + y^3$ &$42$&$ 0 $\\
$13$&$ -y^2 + y^3             $    &$28$&$ -y^2 + y^3 $ &$43$&$ -y + y^2 $\\
$14$&$ -y^2 + y^3 $ &$29$&$ -y + y^2 $ &$44$&$ -1 + y $\\\hline
\hline
\multicolumn{6}{|c|}{$\chi_y(E)=3y(1+y)(1-y)$}\\ \hline
    \end{tabular}}
\end{center}

\begin{center}
Table 6\\*[2mm]
{\small
\begin{tabular}{|r|c|r|c|r|c|}\hline
\multicolumn{6}{|c|}{$({\rm w}_i;d_1)=(n_a;m)=(4,4,5,5,7;25)$}\\ \hline
$\alpha$&$\chi_y^{(\alpha)}$
&$\alpha$&$\chi_y^{(\alpha)}$
&$\alpha$&$\chi_y^{(\alpha)}$\\ \hline
$0$&$  1 + 19y - 19y^3 - y^4 $&$ 9$&$ -y + y^2     $&$18$&$ - y + y^2     $ \\
$1$&$  -y^3 + y^4            $&$10$&$ - 4y + 4y^2  $&$19$&$ - y^2 + y^3   $ \\
$2$&$  -y^2 + y^3            $&$11$&$ - y^2 + y^3  $&$20$&$ - 4y^2 + 4y^3 $ \\
$3$&$  -y + y^2              $&$12$&$ - y  + y^2   $&$21$&$ -y^2 + y^3    $ \\
$4$&$  - y + y^2             $&$13$&$ - y^2 + y^3  $&$22$&$ - y^2 + y^3   $ \\
$5$&$  - 4y + 4y^2           $&$14$&$ -y + y^2     $&$23$&$ - y + y^2     $ \\
$6$&$   - y + y^2            $&$15$&$ - 4y^2 + 4y^3 $&$24$&$ -1 + y       $ \\
$7$&$   - y^2  + y^3         $&$16$&$ - y^2 + y^3   $&$  $&$              $ \\
$8$&$ - y^2 + y^3 $&$17$&$ -y + y^2 $&$ $&$ $ \\\hline
\hline
\multicolumn{6}{|c|}{$\chi_y(E)=3y(1+y)(1-y)$}\\ \hline
    \end{tabular}}
\end{center}
\section{Remark}
While this work was completed, there appeared on the hep-th net a paper of
Kachru \cite{KA}, where some three generation (0,2)
superconformal models with  gauge groups $E_6$ and $SU(5)$ had been found.

\end{document}